\documentclass[aps,prd,preprint,tightenlines,groupedaddress,showpacs]{revtex4}
\usepackage{epsf,epsfig,graphics,graphicx}
\begin{document}

\title{Trapping effects on inflation}

\author{Wolung Lee$^1$}
\author{Kin-Wang Ng$^2$}
\author{I-Chin Wang$^1$}
\author{Chun-Hsien Wu$^3$}
\affiliation{
$^1$Department of Physics, National Taiwan Normal University,
Taipei 116, Taiwan.\\
$^2$Institute of Physics, Academia Sinica, Taipei 115, Taiwan.\\
$^3$Department of Physics, Soochow University,
Taipei 111, Taiwan.}

\date{\today}

\begin{abstract}
We develop a Lagrangian approach based on the influence functional
method so as to derive self-consistently the Langevin equation for
the inflaton field in the presence of trapping points along the
inflaton trajectory. The Langevin equation exhibits the
backreaction and the fluctuation-dissipation relation of the
trapping. The fluctuation is induced by a multiplicative colored
noise that can be identified as the the particle number density
fluctuations and the dissipation is a new effect that may play a
role in the trapping with a strong coupling. In the weak coupling
regime, we calculate the power spectrum of the noise-driven
inflaton fluctuations for a single trapping point and studied its
variation with the trapping location. We also consider a case with
closely spaced trapping points and find that the resulting power
spectrum is blue.
\end{abstract}

\maketitle

\section{Introduction}

The inflationary scenario~\cite{olive}, in which the present
Universe is only a small local patch of a causally connected
region at early times which underwent an exponential expansion
driven by the inflaton potential, is generally accepted for
explaining the observed spatially flat and homogeneous Universe.
In addition, its quantum fluctuations during inflation give rise
to primordial Gaussian matter density fluctuations with a nearly
scale-invariant power spectrum, which is consistent with recent
astrophysical and cosmological observations such as structure
formation and cosmic microwave background
anisotropies~\cite{texas04}.

Although the simplest single-field, slow-roll inflation model
works well, some basic questions have yet to be answered. What is
the origin of the inflaton potential? Do classical matter density
imhomogenities that we observe today genuinely come from quantum
fluctuations of the inflaton? Are the observed matter density
fluctuations truly Gaussian? How robust are the predictions for a
subdominant contribution of tensor modes to the metric
fluctuations, a slightly broken scale invariance, and a negligible
running spectral index of the power spectrum? Future cosmic
microwave background measurements and mega-scale mappings of the
large scale structure will definitely answer some of these
questions or perhaps pose a challenge to the standard inflation
scenario.

There has been a lot of studies on inflationary models that go
beyond the simplest single-field, slow-roll inflation. A class of
models has considered a new source for generating inflaton
fluctuations during inflation through a Yukawa-type or
gravitational interaction between the inflaton and other quantum
fields. This leads to very interesting results such as the
so-called warm inflation~\cite{berera}, the suppression of
large-scale density fluctuations~\cite{wu}, the bursts of particle
production via the infra-red cascading mechanism~\cite{barnaby},
the trapped inflation in which the inflaton rolls slowly down a
steep potential by dumping its kinetic energy into particle
production~\cite{Green,sorbo}, and possible constraints on the
duration of inflationary expansion~\cite{wood}.

In Refs.~\cite{Green,barnaby},the authors analyzed a model in
which the inflaton $\Phi$ couples to another scalar field
$\chi$ via the interaction,
\begin{equation}
\frac{g^{2}}{2}(\Phi-\Phi_{0})^{2}\chi^{2},
\label{interact}
\end{equation}
where $g$ is a coupling constant and $\Phi_{0}$ is a
constant field value. When $\Phi$ rolls down to
$\Phi_{0}$, the $\chi$ particles become instantaneously
massless and are produced with a number density that increases
with $\Phi$'s velocity. As $\Phi$ dumps its
kinetic energy into the $\chi$ particles, it is slowed down and
the produced $\chi$ particles are diluted due to the inflationary
expansion. As shown in Ref.~\cite{Green}, a viable inflationary
model (the so-called trapped inflation) can be achieved by
assuming sufficiently closely spaced trapping points like
$\Phi_{0}$ along the $\Phi$ trajectory even on a
potential which is too steep for slow-roll inflation. In their
approach, the backreaction of the $\chi$ particle production to
the inflaton field and the associated effect of particle number
density fluctuations in the process of particle production are
simply introduced in the equation of motion for $\Phi$ by
use of the mean field approximation:
\begin{equation}
-\nabla_\mu\nabla^\mu\Phi + V'(\Phi)
+g^{2}\langle\chi^{2}\rangle(\Phi-\Phi_0)
=g^{2}\left(\langle\chi^{2}\rangle-\chi^2\right)(\Phi-\Phi_0),
\label{hartree}
\end{equation}
where $V(\Phi)$ is the inflaton potential and we
will go back to this equation later on.

In this paper, we will develop a Lagrangian approach based on the
influence functional method in which the $\chi$ field is
integrated out. This will enable us to derive self-consistently
the equation of motion for the $\Phi$ field that incorporates the
backreaction of the $\chi$ particle production and the particle
number density fluctuations in the particle production. We will
find that the particle number density fluctuations should be
replaced by a colored noise term. In addition, there exists a new
dissipative effect that is closely related to the noise. In fact,
this dissipation is well-known in the real-time approach to
non-equilibrium phenomena in quantum field
theory~\cite{fey,noise}. The paper is organized as follows. In
Sec.~\ref{ifm}, the influence functional method is introduced and
the Langevin equation for the inflaton field is derived. In
Sec.~\ref{trapsec}, we will calculate the power spectrum of the
inflaton fluctuations driven by the noise term in the weak
coupling limit. Also, we will consider the effect of multiple
trapping points on inflaton fluctuations. The dissipation is
discussed in Sec.~\ref{dissipation} and we conclude in
Sec.~\ref{con}.

\section{Influence functional approach}
\label{ifm}

We will adopt the influence functional method~\cite{fey} to take
into account the effects of the quantum fluctuations of the $\chi$
field on the inflaton dynamics in a real-time manner, different from
using a one-loop effective potential as usually found for example
in Ref.~\cite{ala}. Thus, the effective Langevin equation of the
inflaton is obtained, describing the time-dependent corrections to
the inflaton equation of motion originally given by the
classical inflaton potential. In particular, this Langevin equation,
which goes beyond the mean field approximation, involves a
stochastic noise term that drives the growth of perturbation of
the inflaton as we will see in the following.

\subsection{Langevin equation for inflaton}

Let us consider a slow-rolling inflaton $\Phi$ coupled to a
massless scalar field $\chi$ with an interaction given in Eq.(\ref{interact}).
Then, the total Lagrangian is given by
\begin{equation}
\mathcal{L}=\frac{1}{2}g^{\mu\nu}\partial_{\mu}\Phi\,\partial_{\nu}\Phi+
            \frac{1}{2}g^{\mu\nu}\partial_{\mu}\chi\,\partial_{\nu}\chi
             -V(\Phi)- \frac{g^{2}}{2}(\Phi-\Phi_{0})^{2}\chi^{2},
\end{equation}
where $V(\Phi)$ is the inflaton potential that complies with
the slow-roll conditions. To simplify matter, we make a shift:
$\phi=\Phi-\Phi_{0}$. Hence, the Lagrangian becomes
\begin{equation}
\mathcal{L}=\frac{1}{2}g^{\mu\nu}\partial_{\mu}\phi\,\partial_{\nu}\phi+
            \frac{1}{2}g^{\mu\nu}\partial_{\mu}\chi\,\partial_{\nu}\chi
             -V(\phi)- \frac{g^{2}}{2}\phi^{2}\chi^{2},
\label{model}
\end{equation}
where the trapping point is located at $\phi=\phi_0=0$.
We can approximate the space-time during inflation by a de Sitter
metric given by
\begin{equation}
ds^{2}=a^2 (\eta)(d\eta^{2}-d{\bf x}^2),
\end{equation}
where $\eta$ is the conformal time and $a(\eta)= -1/(H\eta)$ with
$H$ being the Hubble parameter. Here we rescale $a=1$ at the initial
time of the inflation era, $\eta_i= -1/H$.

To proceed, let us assume that the initial density matrix at time
$\eta_i$ can be factorized as
\begin{equation}\label{initialcond}
    \rho(\eta_i)=\rho_{\phi}(\eta_i)\otimes\rho_{\chi }(\eta_i)\,.
\end{equation}
The full density matrix evolves unitarily and the evolution can be
described by employing the closed-time-path formalism. Following the
influence functional approach~\cite{fey,noise}, we trace  out the
field $\chi$ in the perturbative expansion. The reduced density
matrix of the system then becomes
\begin{equation}
\rho_{r}(\phi_{f},\phi'_{f};\eta_{f})=\int d\phi_{i} \, d\phi'_{i}
\, \mathcal{Z} (\phi_{f}, \phi'_{f}, \eta_{f} ; \phi_{i}, \phi'_{i},
\eta_{i} ) \,\rho_{r}(\phi_{i},\phi'_{i};\eta_{i}) \, .
\end{equation}
Here the propagating function $\mathcal{Z} (\phi_{f}, \phi'_{f},
\eta_{f} ; \phi_{i}, \phi'_{i}, \eta_{i} )$ is obtained as
\begin{eqnarray}
\mathcal{Z} (\phi_{f}, \phi'_{f}, \eta_{f} ; \phi_{i}, \phi'_{i},
\eta_{i} )& = & \int_{\phi_{i}}^{\phi_{f}} \, \mathcal{D}\phi^+
 \int_{\phi'_{i}}^{\phi'_{f}} \, \mathcal{D}\phi^- \,
e^{i\left( S_0 [\phi^+]-S_0 [\phi^-] \right)} \times e^{i
S_{IF}[\phi^+,J^+,\phi^-,J^-]}\nonumber \\
& = & \mathcal{Z}_{\phi} \cdot \mathcal{Z}_{\chi}  \, ,
\label{propagatingfun}
\end{eqnarray}
with
\begin{equation}
\mathcal{Z}_{\chi}=e^{i S_{IF}[\phi^+,J^+,\phi^-,J^-]}=\int_{\chi_{i}}^{\chi_{f}} \,
\mathcal{D}\chi^+
 \int_{\chi'_{i}}^{\chi'_{f}} \, \mathcal{D}\chi^- \,
e^{i\left( S_0 [\chi^+]-S_0 [\chi^-] \right)} \times e^{i
 \frac{g^2}{2}\int d^4 x \, a^4 (\phi^{+^2} \chi^{+^2}-\phi^{-^2} \chi^{-^2})} \, ,
 \label{influenfun}
\end{equation}
where the actions for the fields $\phi$ and $\chi$ are given by, respectively,
\begin{eqnarray}
S_0 [ \phi ]&=& \int  d^4 x \, a^2 (\eta) \left[ \, \frac{1}{2}
\left(\frac{d\phi}{d\eta}\right)^2 -\frac{1}{2} \left( \nabla \phi
\right)^2 - a^2(\eta) \, V(\phi) \right] \, , \\
S_0 [ \chi ]&=& \int  d^4 x \, a^2 (\eta) \left[ \, \frac{1}{2}
\left(\frac{d\chi}{d\eta}\right)^2 -\frac{1}{2} \left( \nabla \chi
\right)^2 \right] \, ,
\end{eqnarray}
and the sources are $J^+ =\phi^{+^2}$ and $J^- =\phi^{-^2}$.  We can
expand $e^{i S_{IF}}$ in terms of $J^+$ and $J^-$ up to $g^4$ order:
\begin{eqnarray}
S_{IF}[\phi^+,J^+,\phi^-,J^-] &=&
S_{IF}[\phi^+,0,\phi^-,0]+ \int d^4 x  a^4(x) \left[\frac{\delta
S_{IF}}{\delta J^+} J^+  +   \frac{\delta
S_{IF}}{\delta J^-} J^- \right] \nonumber \\
& &+ \frac{1}{2!}  \int\!\!\!\int a^4(x)d^4  x a^4(x') d^4
x^{'} \left[\frac{\delta^2 S_{IF}}{\delta J^{+}_{x} \delta
J^{+}_{x^{'}}}J^{+}_{x} J^{+}_{x^{'}}+  \frac{\delta^2 S_{IF}}{\delta J^{+}_{x} \delta
J^{-}_{x^{'}}}J^{+}_{x} J^{-}_{x^{'}} \right.  \nonumber \\
& &+  \left.
\frac{\delta^2 S_{IF}}{\delta J^{-}_{x} \delta
J^{+}_{x^{'}}}J^{-}_{x} J^{+}_{x^{'}}+  \frac{\delta^2 S_{IF}}{\delta J^{-}_{x}
\delta
J^{-}_{x^{'}}}J^{-}_{x} J^{-}_{x^{'}} \right]+ \, ...\,\,\,\,\,\,,
\label{expansion}
\end{eqnarray}
where the variations with respective to $J^{+}$ and $J^{-}$ are taken at $J^{+}=J^{-}=0$.
The first term $S_{IF}[\phi^+,0,\phi^-,0]$ can be absorbed into the
normalization constant and the remaining terms are given by
\begin{equation}
\frac{\delta S_{IF}}{\delta J^+}=\left[\frac{-i}{\mathcal{Z}_{\chi}}
\frac{\delta \mathcal{Z}_{\chi}}{\delta J^+}\right]_{J^+=0}=\frac{g^2}{2}\langle
\chi^{+^2}\rangle\, , \,\,\,\frac{\delta S_{IF}}{\delta
J^-}=\left[\frac{-i}{\mathcal{Z}_{\chi}} \frac{\delta
\mathcal{Z}_{\chi}}{\delta J^-}\right]_{J^-=0}=\frac{-g^2}{2}\langle \chi^{-^2}
\rangle\, ,\label{firstorder}
\end{equation}
\begin{eqnarray}
\frac{\delta^2 S_{IF}}{\delta J^+ \delta
J^+}&=&-i\left[\frac{-1}{\mathcal{Z}_{\chi}^{2}} \frac{\delta
\mathcal{Z}_{\chi}}{\delta J^+} \frac{\delta
\mathcal{Z}_{\chi}}{\delta J^+} +\frac{1}{\mathcal{Z}_{\chi}}
\frac{\delta^2 \mathcal{Z}_{\chi}}{\delta J^+ \delta
J^+}\right]_{J^+ =0}\nonumber \\
& = &
-i\frac{g^4}{4}\langle\chi^{+^2}(x)\rangle\langle\chi^{+^2}(x^{'})\rangle+i\frac{g^4}{4}\langle\chi^{+^2}(x)
\chi^{+^2}(x^{'})\rangle_{dis} \, ,\label{++}
\end{eqnarray}
\begin{eqnarray}
\frac{\delta^2 S_{IF}}{\delta J^+ \delta
J^-}&=&-i\left[\frac{-1}{\mathcal{Z}_{\chi}^{2}} \frac{\delta
\mathcal{Z}_{\chi}}{\delta J^+} \frac{\delta
\mathcal{Z}_{\chi}}{\delta J^-} +\frac{1}{\mathcal{Z}_{\chi}}
\frac{\delta^2 \mathcal{Z}_{\chi}}{\delta J^+ \delta
J^-}\right]_{J^+ =J^- =0}\nonumber \\
& = &
i\frac{g^4}{4}\langle\chi^{+^2}(x)\rangle\langle\chi^{-^2}(x^{'})\rangle-i\frac{g^4}{4}\langle\chi^{+^2}(x)
\chi^{-^2}(x^{'})\rangle_{dis}\, ,\label{+-}
\end{eqnarray}
\begin{eqnarray}
\frac{\delta^2 S_{IF}}{\delta J^- \delta
J^+}&=&-i\left[\frac{-1}{\mathcal{Z}_{\chi}^{2}} \frac{\delta
\mathcal{Z}_{\chi}}{\delta J^-} \frac{\delta
\mathcal{Z}_{\chi}}{\delta J^+} +\frac{1}{\mathcal{Z}_{\chi}}
\frac{\delta^2 \mathcal{Z}_{\chi}}{\delta J^- \delta
J^+}\right]_{J^+ =J^- =0}\nonumber \\
& = &
i\frac{g^4}{4}\langle\chi^{-^2}(x)\rangle\langle\chi^{+^2}(x^{'})\rangle-i\frac{g^4}{4}\langle\chi^{-^2}(x)
\chi^{+^2}(x^{'})\rangle_{dis}\, ,\label{-+}
\end{eqnarray}
\begin{eqnarray}
\frac{\delta^2 S_{IF}}{\delta J^- \delta
J^-}&=&-i\left[\frac{-1}{\mathcal{Z}_{\chi}^{2}} \frac{\delta
\mathcal{Z}_{\chi}}{\delta J^-} \frac{\delta
\mathcal{Z}_{\chi}}{\delta J^-} +\frac{1}{\mathcal{Z}_{\chi}}
\frac{\delta^2 \mathcal{Z}_{\chi}}{\delta J^- \delta
J^-}\right]_{J^- =0}\nonumber \\
& = &
-i\frac{g^4}{4}\langle\chi^{-^2}(x)\rangle\langle\chi^{-^2}(x^{'})\rangle+i\frac{g^4}{4}\langle\chi^{-^2}(x)
\chi^{-^2}(x^{'})\rangle_{dis}\, , \label{--}
\end{eqnarray}
where the subscript $dis$ means disconnected diagrams. The disconnected
diagrams can be contracted to the connected diagrams as the following:
\begin{equation}
\langle\chi^{+^2}(x)
\chi^{+^2}(x^{'})\rangle_{dis}=2\langle\chi^{+}(x)
\chi^{+}(x^{'})\rangle^{2} + \langle\chi^{+^2}(x)\rangle \cdot
\langle\chi^{+^2}(x^{'})\rangle\, , \label{contraction++}
\end{equation}
\begin{equation}
\langle\chi^{+^2}(x)
\chi^{-^2}(x^{'})\rangle_{dis}=2\langle\chi^{+}(x)
\chi^{-}(x^{'})\rangle^{2} + \langle\chi^{+^2}(x)\rangle \cdot
\langle\chi^{-^2}(x^{'})\rangle\, , \label{contraction+-}
\end{equation}
\begin{equation}
\langle\chi^{-^2}(x)
\chi^{+^2}(x^{'})\rangle_{dis}=2\langle\chi^{-}(x)
\chi^{+}(x^{'})\rangle^{2} + \langle\chi^{-^2}(x)\rangle \cdot
\langle\chi^{+^2}(x^{'})\rangle\, ,\label{contraction-+}
\end{equation}
\begin{equation}
\langle\chi^{-^2}(x)
\chi^{-^2}(x^{'})\rangle_{dis}=2\langle\chi^{-}(x)
\chi^{-}(x^{'})\rangle^{2} + \langle\chi^{-^2}(x)\rangle \cdot
\langle\chi^{-^2}(x^{'})\rangle\, . \label{contraction--}
\end{equation}
After contraction, we are able to obtain the influence functional up
to order $ g^4$ as
\begin{eqnarray}
e^{i S_{IF}[\phi^+,\phi^-]}   & = & \exp \left\{ i  \frac{g^2}{2}
\int d^{4} x_{1}  \, a^4 (\eta_1) \, \left[ \phi^{+ 2}(x_{1}) \,
\langle \chi^+ (x_1) \chi^+ (x_1) \rangle - \phi^{- 2}(x_{1}) \,
\langle
\chi^- (x_1) \chi^- (x_1) \rangle \right] \right. \nonumber \\
& &  - \,\frac{g^{4}}{4} \int d^{4} x_{1}  \int
d^{4} x_{2}   \, a^4 (\eta_1) \, a^4(\eta_2) \nonumber \\
&&  \left[ \phi^{+ 2}(x_{1}) \, \langle \chi^+ (x_1) \chi^+ (x_2)
\rangle^2 \, \phi^{+2}(x_{2}) -  \phi^{+ 2}(x_{1}) \, \langle \chi^+
(x_1)
\chi^- (x_2) \rangle^2 \, \phi^{-2}(x_{2}) \right. \nonumber \\
 &  &  \left. \left.-  \phi^{- 2}(x_{1}) \, \langle \chi^- (x_1) \chi^+ (x_2)
\rangle^2 \, \phi^{+2}(x_{2}) +  \phi^{- 2}(x_{1}) \, \langle \chi^-
(x_1) \chi^- (x_2) \rangle^2 \, \phi^{-2}(x_{2}) \right] \right\} \, .
\label{influencefun}
\end{eqnarray}
The Green's functions of the $\chi$ field are defined by
\begin{eqnarray}
    \bigl< \chi^+ (x) \chi^+ (x')\bigr>&=& \bigl< \chi (x) \chi(x')\bigr> \,\theta(\eta-\eta')
    +\bigl< \chi(x')\chi(x) \bigr> \,\theta (\eta'-\eta)\,,\nonumber\\
    \bigl< \chi^- (x) \chi^- (x')\bigr> &=& \bigl< \chi (x') \chi(x)\bigr>\,\theta(\eta-\eta')
    +\bigl< \chi(x)\chi(x')\bigr> \,\theta (\eta'-\eta)\,,\nonumber\\
    \bigl< \chi^+ (x) \chi^- (x')\bigr>&=&\bigl< \chi (x) \chi(x')\bigr>\,,\nonumber\\
    \bigl< \chi^- (x) \chi^+ (x')\bigr>&=& \bigl< \chi(x')\chi(x)\bigr> \,,
\end{eqnarray}
and can be explicitly constructed as long as its vacuum state has
been specified. To obtain the semiclassical Langevin equation, it is
more convenient to introduce the average and relative field
variables:
\begin{equation}
 \phi =\frac{1}{2}( \phi^+ + \phi^- ) \,\,\, , \,\,\,
 \phi_\Delta=\phi^+ - \phi^- \, .
\end{equation}
The coarse-grained effective action (CGEA) including the influence
action $S_{IF}$ obtained from Eqs.~(\ref{propagatingfun}) to~(\ref{influencefun})
is then given by
\begin{eqnarray}
S_{CGEA} \left[ \phi, \phi_{\Delta} \right]&=& \int d^{4} x \,
a^2 (\eta) \, \phi_{\Delta} (x) \left\{ -\ddot{\phi}(x) -
2aH\dot{\phi}(x) +\nabla^{2}\phi(x)  - a^{2}\left[ V'(\phi)
+g^{2}\langle\chi^{2}\rangle\phi(x) \right] \right. \nonumber \\
&& \left.
\,\,\,\,\,\,\,\,\,\,\,\,\,\,\,\,\,\,\,\,\,\,\,\,\,\,\,\,\,\,\,\, -
g^4 a^2 (\eta) \phi (x)  \int d^{4} x' \, a^4(\eta')
\,\theta(\eta-\eta')\, i G_- (x,x') \phi^2 (x') \right\}\,
\nonumber \\
&+&  i \frac{g^4}{2} \int  d^{4} x  \int d^{4} x' a^4 (\eta) a^4
(\eta') \, \phi_{\Delta} (x) \phi (x) \, G_+ ( x,x') \,
\phi_{\Delta} (x') \phi (x') + \mathcal{O} ( \phi_{\Delta}^3) \, ,
\nonumber \\
\end{eqnarray}
where the dot and prime denote respectively differentiation with
respect to $\eta$ and $\phi$. In addition, we have used the fact
that correlation functions of the fields evaluated at the same
space-time point in the $+$ and $-$ branches are equal, namely,
$\langle \chi^+ (x_1) \chi^+ (x_1) \rangle=\langle \chi^- (x_1)
\chi^- (x_1) \rangle \equiv \langle \chi^2 (x_1) \rangle$. The
kernels $G_{\pm}$ can be obtained from the Green's function of
$\chi$:
\begin{equation}
G_{+}(x,x')=\langle\chi(x)\chi(x')\rangle^2 +
                               \langle\chi(x')\chi(x)\rangle^2\, ,
\label{grfct}
\end{equation}
\begin{equation}
G_{-}(x,x')=\langle\chi(x)\chi(x')\rangle^2 -
                               \langle\chi(x')\chi(x)\rangle^2\, .
\label{grfct2}
\end{equation}
The imaginary part of the above influence action can be re-expressed
by introducing an auxiliary field $ \xi $  with a distribution
function of the Gaussian form,
\begin{equation}
P[\xi ] = \exp \left\{ - \frac{1}{2}  \int d^{4}x \, \int d^{4} x'
\, \xi (x) \, \nu^{-1} (x,x') \, \xi (x') \right\} \, ,
\label{noisedistri}
\end{equation}
where the noise kernel is
\begin{equation}
\nu(x,x')=\langle\xi(x) \xi(x')\rangle= G_{+}(x,x')\, .
\end{equation}
This leads to
\begin{equation}
e^{i S_{CGEA}}  =  \int {\cal D} \xi \, P [\xi ] \, \exp i S_{\rm
eff} \left[ \phi , \phi_{\Delta}, \xi \right] \, ,
\end{equation}
with the effective action $ S_{\rm eff}$ given by
\begin{eqnarray}
S_{\rm eff} [\phi,\phi_{\Delta}, \xi ] && = \int d^4 x \, a^2 (\eta)
\, \phi_{\Delta} (x) \left\{ -\ddot{\phi}(x) - 2aH\dot{\phi}(x)
+\nabla^{2}\phi(x)  - a^{2}\left[ V'(\phi)
+g^{2}\langle\chi^{2}\rangle\phi(x) \right] \right. \nonumber \\
&&-\left.   g^4 a^2 \phi (x) \int d^{4} x' \, a^4(\eta') \,
\theta(\eta-\eta') \, i G_- (x,x') \phi^2 (x')    + g^2 a^2 \phi (x)
\xi(x) \right\} \, .
\end{eqnarray}
The semiclassical approximation requires to extremize the effective
action $ \delta S_{\rm eff}/ \delta \phi_{\Delta}$ when long-wavelength
inflaton modes of cosmological interest have gone through the
quantum-to-classical transition due to the rapid expansion of the scale
factor~\cite{sta}. Then, we obtain the semiclassical Langevin
equation for $\phi$:
\begin{eqnarray}
&& \ddot{\phi}+2aH\dot{\phi}-\nabla^{2}\phi+a^{2}\left[V'(\phi)
+g^{2}\langle\chi^{2}\rangle\phi\right]+g^{4}a^{2}{\phi}  \int d^{4}
x'
 a^4(\eta') \times \nonumber \\
&& \,\,\,\,\,\,\,\,\,\,\,\,\,\,\,\,\,\,\,\,\,
\theta(\eta-\eta')\,i\, G_{-}(x,x') {\phi}^{2}(x')=g^2\,a^2\, \phi
\, \xi \, . \label{lange}
\end{eqnarray}
This is the main result of our paper. It shows that the effects
from the quantum field $\chi$ on the inflaton are given by the
dissipation via the kernel $G_{-}$ as well as the fluctuation
induced by the multiplicative colored noise $\xi$ with
\begin{equation}
\langle\xi(x)\xi(x')\rangle=  G_{+}(x,x').
\label{noiseterm}
\end{equation}

\subsection{Approximate solutions}

To solve Eq.~(\ref{lange}), let us first drop the dissipative term
which we will discuss later and consider the colored noise only.
Then, Eq.~(\ref{lange}) becomes
\begin{equation}
\ddot{\phi}+2aH\dot{\phi}-\nabla^{2}\phi+a^{2}\left[V'(\phi)
+g^{2}\langle\chi^{2}\rangle\phi\right]=g^2\,a^2\, \phi
\, \xi \, . \label{lange2}
\end{equation}
After decomposing $\phi$ into a mean field and a classical
perturbation: $\phi(\eta,\bf x)={\bar\phi}(\eta) +
{\varphi}(\eta,\bf x)$, we obtain the linearized Langevin equation,
\begin{equation}
\ddot{\varphi}+2aH\dot{\varphi}-\nabla^{2}{\varphi}+ a^{2}
m_{\varphi{\rm eff}}^2 {\varphi} = g^2\, a^2\, {\bar\phi}\, \xi,
\label{varphieq}
\end{equation}
where the effective mass is $m_{\varphi{\rm
eff}}^2=V''({\bar\phi})+g^{2}\langle\chi^{2}\rangle$ and the time
evolution of $\bar\phi$ is governed by
\begin{equation}
\ddot{\bar\phi}+2aH\dot{\bar\phi}+a^{2}\left[V'(\bar\phi)
+g^{2}\langle\chi^{2}\rangle{\bar\phi}\right]=0.
\label{meanphieq}
\end{equation}
The equation of motion for $\chi$ from which we construct its Green's function
can be read off from its quadratic terms in the
Lagrangian~(\ref{model}) as
\begin{equation}
\ddot{\chi}+2aH\dot{\chi}-\nabla^{2}{\chi}+ a^{2} m_{\chi{\rm eff}}^2 {\chi}
= 0, \label{chieq}
\end{equation}
where the effective mass is $m_{\chi{\rm eff}}^2=g^2{\bar\phi}^2$.
Let us decompose
\begin{eqnarray}
Y(x)&=&\int\frac{d^3{\bf k}}{(2\pi)^{3\over 2}} Y_{\bf
k}(\eta)\,e^{i{\bf k}\cdot{\bf x}}, \quad
{\rm where}\;Y=\varphi,\xi, \nonumber \\
\chi(x)&=&\int\frac{d^3{\bf k}}{(2\pi)^{3\over 2}} \left[b_{\bf
k}\chi_k(\eta)\,e^{i{\bf k}\cdot{\bf x}} + {\rm h.c.}\right],
\end{eqnarray}
where $b_{\bf k}^\dagger$ and $b_{\bf k}$ are creation and
annihilation operators satisfying $[b_{\bf k},b_{{\bf
k}'}^{\dagger}]= \delta({\bf k}-{\bf k}')$.
Then, the solution to Eq.~(\ref{varphieq}) is obtained as
\begin{equation}
\varphi_{\bf k} (\eta)= g^2\int_{\eta_i}^{\eta}
d\eta' a^2(\eta') {\bar\phi}(\eta') \xi_{\bf
k}(\eta') G_r(\eta',\eta),
\label{varphisol}
\end{equation}
where we have adopted the retarded Green's function and
\begin{equation}
G_r(\eta',\eta)=\left[\varphi_k^{1}(\eta')\varphi_k^{2}(\eta)
                   - \varphi_k^{2}(\eta')\varphi_k^{1}(\eta)\right]
                   W^{-1}\left[\varphi_k^{1}(\eta'),\varphi_k^{2}(\eta')\right].
\end{equation}
Here the homogeneous solutions $\varphi_k^{1,2}$ of Eq.~(\ref{varphieq}) are given by
\begin{equation}
\varphi_k^{1,2}={1\over2a} (\pi|\eta|)^{1\over 2}
                        H_\nu^{(1),(2)}(k\eta)
\end{equation}
and the Wronskian is
\begin{equation}
W\left[\varphi_k^{1}(\eta),\varphi_k^{2}(\eta)\right]=
\varphi_k^{1}(\eta)\frac{d\varphi_k^{2}(\eta)}{d\eta}-\varphi_k^{2}(\eta)\frac{d\varphi_k^{1}(\eta)}{d\eta}=
\frac{i}{a^{2}(\eta)}\,.
\end{equation}
$H_\nu^{(1)}$ and $H_\nu^{(2)}$ are Hankel functions of the
first and second kinds respectively and $\nu^2=9/4-m_{\varphi{\rm
eff}}^2/H^2$. In addition, we have from Eq.~(\ref{chieq}) that
\begin{equation}
\chi_{ k} ({\eta})={1\over2a} (\pi|\eta|)^{1\over 2} \left[c_1
H_\mu^{(1)}(k\eta)+c_2 H_\mu^{(2)}(k\eta)\right],
\label{chisolution}
\end{equation}
where the constants $c_1$ and $c_2$ are subject to the normalization
condition, $|c_2|^2 -|c_1|^2=1$, and $\mu^2=9/4-m_{\chi{\rm eff}}^2/H^2$.

\section{Trapping effects to inflation}
\label{trapsec}

The Langevin equation~(\ref{lange2}) that we have derived by going
beyond the mean field approximation, though similar to
the equation of motion~(\ref{hartree}) used in Refs.~\cite{barnaby,Green},
has more physical meanings. While the backreaction is a common feature,
the particle number density fluctuations in Eq.~(\ref{hartree})
are replaced by a colored noise~(\ref{noiseterm}) that
appears as a source term in Eq.~(\ref{lange2}).
In addition, the full Langevin equation~(\ref{lange})
has a new dissipation term that is expected to co-exist with the noise term
by virtue of the fluctuation-dissipation theorem~\cite{noise}.

Now we are ready to calculate the power spectrum of the perturbation
$\varphi$ induced by the noise term. We already have $V''({\bar\phi})\ll H^2$
for a slow-roll inflaton potential. In order to maintain the slow-roll condition:
$m_{\bar\phi{\rm eff}}^2=m_{\varphi{\rm eff}}^2\ll H^2$ (i.e., $\nu=3/2$),
we further require that $g^{2}\langle\chi^{2}\rangle\ll H^2$.

\subsection{Weak coupling limit}

\begin{figure}[htp]
\centering
\includegraphics[width=0.9\textwidth]{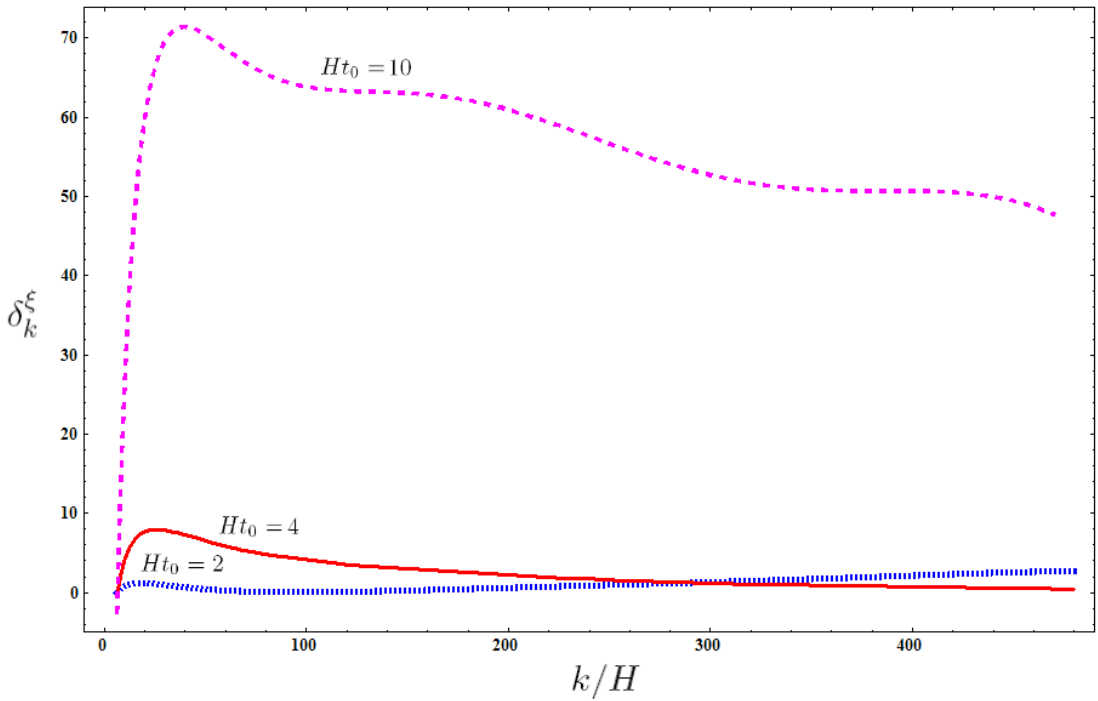}
\caption{Power spectra of the noise-driven inflaton fluctuations
$\delta^\xi_k\equiv 2\pi^2 H^2\Delta^\xi_k/(g^4 v^2)= (2\pi^2/g^4)(2\pi/H)^2 P_\zeta \Delta^\xi_k$,
where $P_\zeta$ is defined in Eq.~(\ref{Pzeta}), with the trapping points
located at $Ht_0=2,4,10$ respectively.
The starting point, $k/H=2\pi$, corresponds to the $k$-mode that leaves
the horizon at the start of inflation.} \label{ps4one}
\end{figure}

We consider, as the simplest case, a very weak coupling constant, $g^2\ll 1$.
This allows us to highlight the effect of trapping to inflation, although
the effect is too small to be observed.
In the weak coupling limit, we simply have $\nu=\mu=3/2$.
It was shown that when $\mu=3/2$ one can select the Bunch-Davies vacuum
(i.e., $c_2=1$ and $c_1=0$) in Eq.~(\ref{chisolution})~\cite{fordkkk}. Hence, using
Eqs.~(\ref{noiseterm}) and (\ref{varphisol}), we obtain
\begin{equation}
\langle\varphi_{\bf k}(\eta)\varphi_{{\bf k}'}^*(\eta)\rangle
=\frac{2\pi^2}{k^3}\Delta^\xi_k(\eta) \delta({\bf k}-{\bf k}'),
\end{equation}
where the noise-driven power spectrum is given by
\begin{equation}
\Delta^\xi_k(\eta)=\frac{g^4 z^2}{8\pi^4} \int_{z_i}^z dz_1
\int_{z_i}^z dz_2 \, {\bar\phi}(\eta_1) {\bar\phi}(\eta_2) F(z_1)
F(z_2)\left\{ \frac{\, \sin z_{-}}{z_1 z_2 z_{-}} \left[\, \sin(2\Lambda
z_{-}/k)/z_{-}-1\right]+G(z_1,z_2)\right\} , \label{pseq}
\end{equation}
where $z_{-}=z_2-z_1$, $z=k\eta$, $z_i=k\eta_i=-k/H$, $\Lambda$ is
the momentum cutoff introduced in the evaluation of the ultraviolet
divergent $k$-integration of $\chi_k$ in the Green's
function~(\ref{grfct}),
\begin{equation}
F(y)=\left(1+\frac{1}{yz}\right)\sin(y-z)+ \left({1\over y}-{1\over
z}\right)\cos(y-z),
\end{equation}
and
\begin{eqnarray}
&&G(z_1,z_2)\nonumber\\
&=& \int_{0}^\Lambda dk_1 \, \int_{\mid k-k_1\mid}^{k+k_1} dq
\left\{\left[\frac{2}{z_1 z_2 q
k_1}\left(\frac{k_1}{q}-\frac{z_1}{z_2}-\frac{z_2}{z_1}+2+\frac{k^2}{z_1
z_2 k_1 q}\right)\right]\, \cos \left[\left(\frac{k_1+q}{k}\right)
(z_2 - z_1)\right]\right. \nonumber
\\&& + \, \left.\frac{2}{kq}\left(\frac{1}{z_1}-\frac{1}{z_2}\right)\left[1+\frac{k^2}{z_1 z_2 k_1}
\left(\frac{1}{q}+\frac{1}{k_1}\right)\right]\,\sin\left[\left(\frac{k_1+q}{k}\right)(z_2-z_1)\right]\right \}\nonumber
\\&& + \frac{2}{z_1 z_2} \int_{0}^\Lambda \frac{dk_1}{k_1}\left\{ \cos(z_2 - z_1) -
\cos\left[(z_2 - z_1)\left(\frac{2k_1}{k}-1\right)\right]\right \}\nonumber
\\&& + \frac{2k}{z_1 z_2 (z_2 - z_1)} \left\{ \int_{0}^\Lambda \frac{dk_1}{k_1^2}
\sin\left[(z_2 - z_1)\left(1+\frac{2k_1}{k}\right)\right]\right. \nonumber
\\&& \left.-\int_{0}^k \frac{dk_1}{k_1^2} \sin(z_2 - z_1) -
\int_{k}^\Lambda \frac{dk_1}{k_1^2}\sin\left[(z_2 - z_1)\left(\frac{2k_1}{k}-1\right)\right]\right\}.
\label{Gintegral}
\end{eqnarray}
Note that the term $\sin(2\Lambda z_{-}/k)/z_{-} \simeq
\pi\delta(z_{-})$ in Eq.~(\ref{pseq}) when $\Lambda\gg k$.
Now we can approximate $\bar\phi(\eta)$ by
\begin{equation}
{\bar\phi}(\eta)=v(t_0-t)=\frac{v}{H}\ln\frac{\eta}{\eta_0},
\label{vel}
\end{equation}
where the real time $t$ is defined by $a(\eta)=a(t)=e^{Ht}$,
$t_0$ is the time when $\bar\phi$ reaches the trapping point,
and $v$ is the slow-roll velocity of $\bar\phi$. Given the inflaton
potential $V(\bar\phi)$, $v\simeq -V'(\bar\phi)/(3H)$.
After substituting Eq.~(\ref{vel}) for ${\bar\phi}(\eta_1)$ and ${\bar\phi}(\eta_2)$
in Eq.~(\ref{pseq}), we numerically evaluate the multiple integral.
In Fig.~\ref{ps4one}, we plot $\Delta^\xi_k(\eta)$ evaluated at the horizon-crossing time,
which is given by $z=-2\pi$, versus $k/H$ for various locations of the trapping point
along the inflaton trajectory. Using the e-folding $Ht_0$ to mark the
moment when $\bar\phi$ hits the trapping point, we investigate the effect of
trapping by putting the trapping point at $Ht_0=2,4,10$. In the figure, the modes
with $k/H=2\pi$ and $k/H=500$ cross out the horizon at the start of inflation and
at about $4.4$ e-folds, respectively.
The figure shows the dependence of the noise-driven fluctuations
on the onset time of inflation, reflecting the integrated effect of
the source powered by the noise term. In the case with $Ht_0=2$,
the multiplicative noise $\bar\phi\xi$ in Eq.~(\ref{varphieq})
vanishes at the trapping point before the mode with $k/H=500$ leaves
the horizon, resulting in a dip in the power spectrum.
When $Ht_0$ increases, the trapping effect becomes insignificant
for the range of $k$-mode in the figure.
In evaluating the integral in Eq.~(\ref{Gintegral}), we have set
the cutoff scale $\Lambda=200\pi$ to obtain the power spectra
and found that the results are insensitive to the choice of
the cutoff value.

\subsection{Closely spaced trapping points}

It is useful to define a time scale, $\Delta t \equiv 1/\sqrt{gv}$.
Then, We have
\begin{equation}
H\Delta t=\left(\frac{2\pi}{g}\right)^{1\over2} P_\zeta^{1\over4},
\end{equation}
where $P_\zeta$ is the matter power spectrum given by
\begin{equation}
P_\zeta=\left(\frac{H}{v}\right)^2 \left(\frac{H}{2\pi}\right)^2.
\label{Pzeta}
\end{equation}
It has been measured by the seven-year WMAP to be
$P_\zeta \simeq 2.4\times 10^{-9}$~\cite{wmap7}.

The $\chi$ particle production can be obtained by solving Eq.~(\ref{chieq}),
in which the effective mass is
\begin{equation}
m_{\chi{\rm eff}}^2=g^2{\bar\phi}^2=g^2 v^2(t_0-t)^2=\Delta t^{-4}(t_0-t)^2.
\end{equation}
When $t\simeq t_0$, the $\chi$ particles become instantaneously
massless and particle production begins. For $H\Delta t < 1$ or $g^2>10^{-7}$,
it was shown~\cite{barnaby,Green} that bursts of particle production takes place
in a time scale, $\Delta t$, and the number density of
the $\chi$ particles produced is given by $n_\chi \simeq 1/\Delta t^3$.
This leads to the backreaction of the $\chi$ particle production to
the inflaton field and the associated effect of particle number
density fluctuations in the process of particle production.
Furthermore, for closely spaced trapping points with equal spacing $\Gamma$
along the inflaton trajectory, trapped inflation occurs even on a
potential which is too steep for slow-roll inflation, provided that
$H\Gamma/v\ll 1$~\cite{Green}.

Here we consider the case for $H\Delta t \simeq 1$ or $g^2\simeq 10^{-7}$.
It is well-known~\cite{fordkkk} that for a massless field,
the quantum fluctuations grow linearly with time:
$\langle\chi^{2}\rangle\simeq H^3 t/(4\pi^2)$. As $\bar\phi$ is moving
away from the trapping point, $\chi$ acquires an effective mass and the
growth slows down. This happens in a time scale of $\Delta t$, so
$n_\chi=m_{\chi{\rm eff}}\langle\chi^{2}\rangle/2=H^3/(8\pi^2)$.
This is consistent with the estimate in the above paragraph.
Also, $g^2\langle\chi^{2}\rangle\simeq 10^{-9}H^2$,
which satisfies the slow-roll requirement. Therefore, within the time
scale $\Delta t$, both $m_{\chi{\rm eff}}^2$ and $m_{\phi{\rm eff}}^2$
are smaller than $H^2$ and we can approximate the homogeneous $\varphi$ solution
and the $\chi$ solution by the mode functions in Eq.~(\ref{varphisol})
with $\nu=3/2$ and Eq.~(\ref{chisolution}) with $\mu=3/2$, respectively.

\begin{figure}[htp]
\centering
\includegraphics[width=0.9\textwidth]{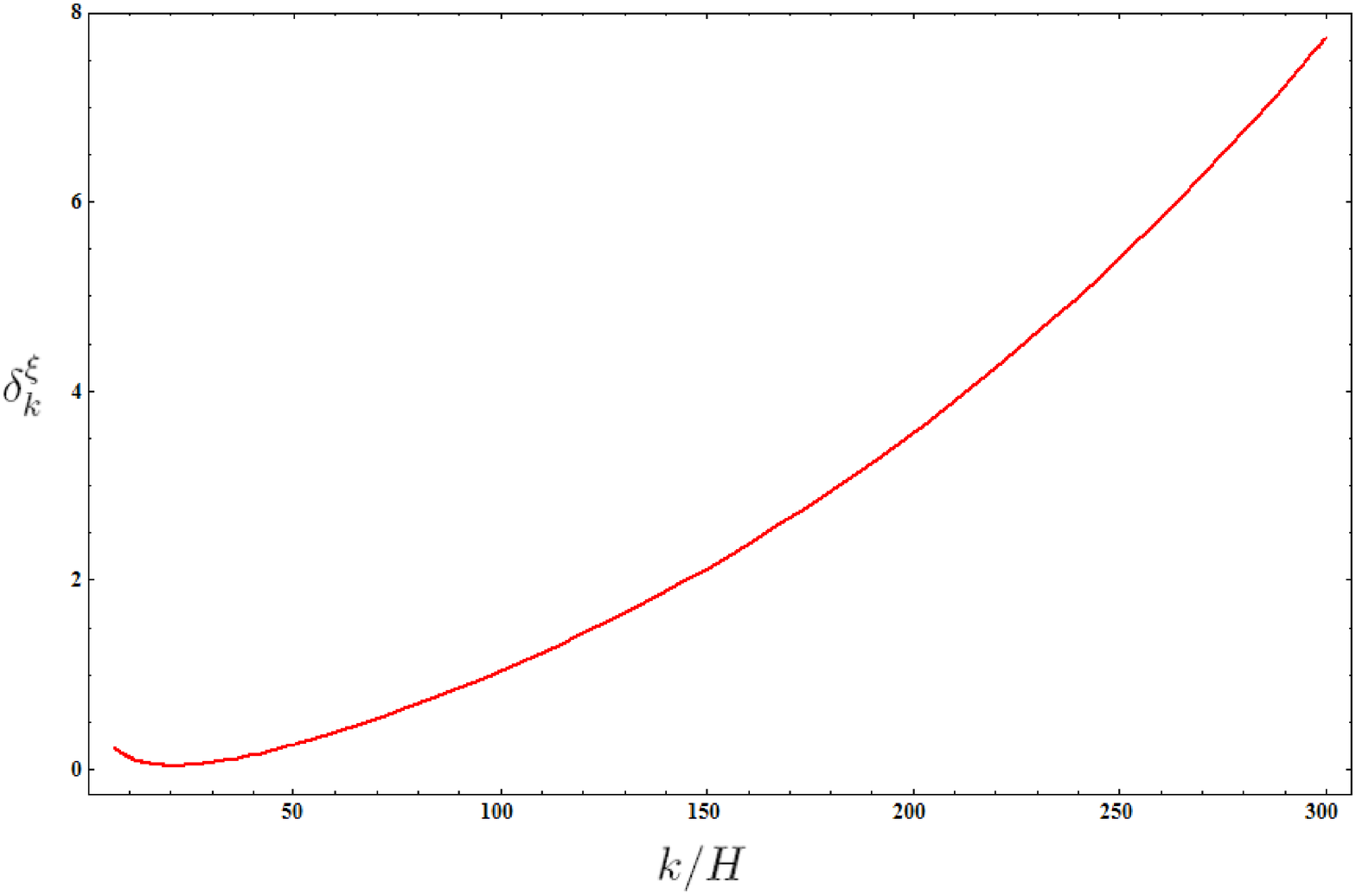}
\caption{Accumulative power spectrum of the noise-driven inflaton fluctuations
$\delta^\xi_k\equiv 10^{-5}\,8\pi^4 v^2 \Delta^\xi_k/(g^4 \Gamma^4 H^2)$
for the case with closely spaced trapping points with even spacing
$\Gamma$ along the inflaton trajectory.}
\label{ps4many}
\end{figure}

Assume that there exist evenly spaced trapping points with $\Gamma/v < \Delta t$
along the inflaton trajectory. For each trapping event,
${\bar\phi}(\eta')$ in Eq.~(\ref{varphisol}) can be approximately
replaced by the constant spacing $\Gamma$ and
\begin{equation}
G_r(\eta',\eta)=G_r(\eta',\eta'+\Delta\eta')\simeq \Delta\eta'=\frac{\Gamma}{v}\frac{1}{a(\eta')}\,.
\end{equation}
Then, we can estimate the noise-driven inflaton fluctuations in this single trapping as
\begin{equation}
d\varphi_{\bf k} (\eta')\simeq \frac{g^2\Gamma^2}{v}
d\eta' a(\eta') \xi_{\bf k}(\eta').
\end{equation}
Hence, at time $\eta$ the accumulative noise-driven inflaton fluctuations are given by
\begin{equation}
\varphi_{\bf k} (\eta)\simeq \frac{g^2\Gamma^2}{v}
\int_{\eta_i}^{\eta} d\eta' a(\eta') \xi_{\bf k}(\eta'),
\end{equation}
and the power spectrum is
\begin{equation}
\Delta^\xi_k(\eta)=\frac{g^4 \Gamma^4 H^2}{8\pi^4 v^2} \int_{z_i}^z dz_1
\int_{z_i}^z dz_2 \, z_1 z_2\left\{ \frac{\, \sin z_{-}}{z_{-}} \left[ \sin(2\Lambda
z_{-}/k)/z_{-}-1\right]+G(z_1,z_2)\right\}.
\end{equation}
We plot this $\Delta^\xi_k(\eta)$ at the horizon-crossing
time given by $z=-2\pi$ versus $k/H$ in Fig.~\ref{ps4many}.
The figure shows that the total noise-driven fluctuations
due to closely spaced trapping points depend on the onset time
of inflation and have a blue power spectrum $\Delta^\xi_k$ that increases with $k$.
When $\Gamma$ saturates the upper limit, i.e.
$\Gamma=v\Delta t$, we have $\Gamma=1/(g\Delta t)\simeq H/g$ and
$\Delta^\xi_k$ increases to about $10^6 g^2 H^2/(8\pi^4)$ at $k/H=300$,
which is still much smaller than the intrinsic de Sitter quantum
fluctuations by a factor of about $4\pi^2/(10^6 g^2)\simeq 400$.

\section{Dissipation term}
\label{dissipation}

The dissipation term in the Langevin equation~(\ref{lange})
for inflaton is of order $g^4$.
In Ref.~\cite{wu}, a model that has the Lagrangian~(\ref{model})
plus a mass term $H^2\chi^2/2$ for the $\chi$ field was considered.
It was shown that including the dissipation term would only slightly affect
the kinematics for slow-roll inflation as well as the noise-driven inflaton fluctuations
even with $g^2\sim 1$. In this paper, we have considered a weak coupling constant
with $g^2< 10^{-7}$. Although the $\chi$ field is massless here, the results
for the noise-driven inflaton fluctuations are similar to those in Ref.~\cite{wu}.
Therefore, we expect that the dissipation term can be safely omitted.
However, for strong couplings with $g^2>10^{-7}$, such as those considered
in the study of particle bursts~\cite{barnaby} and in the trapped inflation~\cite{Green},
the dissipation term may be in competition with the noise term. If so, one
would need to solve the full Langevin equation~(\ref{lange})
to re-examine the inflation dynamics and the inflaton fluctuations.

\section{Conclusions}
\label{con}

We have developed a Lagrangian approach based on the
influence functional method to investigate the effects of trapping
on inflation. The Langevin equation for inflaton thus derived in
Eq.~(\ref{lange}) is compared to the equation of motion used in
Refs.~\cite{Green,barnaby}. The multiplicative colored noise in
the Langevin equation is indeed the particle number density fluctuations
studied by the authors in Refs.~\cite{Green,barnaby}; however, they
have not considered the dissipation term.

The Langevin equation has been solved in the weak coupling regime
with $g^2< 10^{-7}$, in which the dissipation can be ignored.
We have calculated the power spectrum of the noise-driven inflaton fluctuations
for a single trapping point and studied its variation with the location of
the trapping point along the inflaton trajectory. We have found that
if the inflaton rolls down the potential with closely spaced trapping points,
the resulting power spectrum would be blue. This is an interesting result. However,
the dissipation should begin to play a role and damp the power of the
high $k$ modes which leave the horizon in late times. 

The present paper has given a systematic approach to deal with
trapping points located along the inflaton trajectory.
It worths re-examining the effects of trapping with a strong
coupling on inflation that have been discussed in
Refs.~\cite{Green,barnaby}. In particular, it is interesting
to study the effect of dissipation to both the backreaction
and the fluctuations.

\begin{acknowledgments}

We would like to thank B.-L. Hu and E. Silverstein for useful
discussions. This work was supported in part by the National
Science Council, Taiwan, ROC under the Grants
NSC97-2112-M-003-004-MY3 (W.L.L.), NSC98-2112-M-001-009-MY3
(K.W.N.), and NSC99-2112-M-031-002-MY3 (C.H.W.).

\end{acknowledgments}

\end{document}